\documentclass[a4paper,english,aps,pre,showpacs,twocolumn,a4paper,floatfix]{revtex4}
\usepackage[T1]{fontenc}
\usepackage[latin1]{inputenc}
\usepackage{graphicx}
\usepackage{epsfig}
\begin{document}
\title{Enhanced stability of tetratic phase due to clustering}

\author{Yuri Mart\'{\i}nez-Rat\'on}
\email{yuri@math.uc3m.es}

\address{Grupo Interdisciplinar de Sistemas Complejos (GISC), Departamento
de Matem\'aticas, Escuela
Polit\'ecnica Superior, Universidad Carlos III de Madrid, Avenida de la
Universidad 30, 28911-Legan\'es, Madrid, Spain}

\author{Enrique Velasco}
\email{enrique.velasco@uam.es}

\address{Departamento de F\'{\i}sica Te\'orica de la Materia Condensada and 
Instituto de Ciencia
de Materiales Nicol\'as Cabrera, Universidad Aut\'onoma de Madrid, E-28049 
Madrid, Spain}
\begin{abstract}
We show that the relative stability of the nematic tetratic phase with
respect to the usual uniaxial nematic phase can be greatly enhanced by 
clustering effects. Two--dimensional rectangles of aspect ratio $\kappa$
interacting via hard interactions are considered, and the stability
of the two nematic phases (uniaxial and tetratic) is examined using 
an extended scaled--particle theory applied to a polydispersed fluid
mixture of $n$ species. Here the $i$--th species is associated with clusters 
of $i$ rectangles, with clusters defined as stacks of rectangles containing 
approximately parallel rectangles, with 
frozen internal degrees of freedom. The theory assumes an exponential cluster 
size distribution (an assumption fully supported by Monte Carlo simulations
and by a simple chemical--reaction model),
with fixed value of the second moment. The corresponding area distribution 
presents a shoulder, and sometimes even a well-defined peak,
at cluster sizes approximately corresponding to square shape (i.e. $i\simeq\kappa$),
meaning that square clusters have a dominant contribution to the free energy of the
hard--rectangle fluid. The theory predicts an enhanced region of stability of the 
tetratic phase with respect to the standard scaled--particle theory, much closer 
to simulation and to experimental results, demonstrating the importance of
clustering in this fluid.
\end{abstract}

\date{\today}
\maketitle

\section{Introduction}
The hard rectangle (HR) fluid constitutes a paradigmatic example of a two-dimensional 
fluid exhibiting surprisingly complex phase behavior: different phase symmetries, 
phase transitions with different order, and defect--mediated continuous transitions
of the Kosterlitz--Thouless type \cite{Chaikin}, all governed solely by entropy.
This peculiar two--dimensional system has three equilibrium fluid phases:
isotropic (I), where particle axes are randomly oriented, 
uniaxial nematic (N$_{\rm{u}}$), with particles preferentially aligned along
a single nematic director, and tetratic nematic (N$_{\rm{t}}$), possessing
two equivalent
perpendicular nematic directors, with long particle axes oriented along one of  
two directors with equal probability. 

In a pioneering study, Schlacken et al. \cite{Schlacken} applied scaled--particle theory 
(SPT) on a fluid of HRs to demonstrate the stability of the N$_{\rm{t}}$ phase,
a phase which cannot be stabilised in a fluid of hard ellipses.  
The intersection between the two spinodals associated with the I-N$_{\rm{t}}$ and 
I-N$_{\rm{u}}$ transitions defines a limiting aspect ratio $\kappa=L/\sigma_0$ 
(with $L$ and $\sigma_0$ the length and width of the rectangles, respectively)
for the stability of the N$_{\rm t}$ phase, which is located at $\kappa\simeq 2.62$. 
Thus, for lower values of aspect ratio, the isotropic fluid exhibits 
a continuous transition to the N$_{\rm{t}}$ phase, whereas if $\kappa>2.62$ the I phase
goes directly to the uniaxial nematic phase N$_{\rm u}$.

The study of Schlacken et al. was later supplemented by the calculation of the 
complete phase diagram \cite{Yuri1}, also within the context of SPT. It was found, 
in particular, that the N$_{\rm{t}}$ phase undergoes a transition to the N$_{\rm{u}}$ phase
at high density, the nature of which changes from second to first order at a tricritical point.
In addition, it was shown \cite{Yuri1} that, at the level of a particular approximation for
density--functional theory, the N$_{\rm{t}}$ fluid is metastable with respect to a phase
with (either partial or complete) spatial order; the theory was approximate in the 
sense that it included the exact functional form of two-body correlations, but only 
approximate higher--order correlations. On the other hand, Monte Carlo (MC) simulations 
conducted on hard squares \cite{Frenkel} and on a HR system \cite{Donev} with $\kappa=2$
indicated, as expected, that the fluid exhibits quasi--long--range tetratic order, and 
that the high--density phase consists of an aperiodic crystalline tetratic phase 
exhibiting random tiling on a square lattice. MC simulation of a HR fluid confined
in a slit pore \cite{Fichthorn} show the presence of weak tetratic correlations
at the centre of the pore.

In the experimental front, recent results for colloidal discs forced to stand on edge
by external potentials \cite{Chaikin} (and hence interacting approximately as HRs) have also 
demonstrated that tetratic correlations play a vital role in this system. Also,
experiments conducted on a monolayer of vibrated granular cylinders lying on a plate
\cite{Narayan} have shown tetratic correlations for cylinders with aspect ratio as high 
as $\kappa=12.6$.

In Ref. \cite{Yuri2} strong evidence, based on MC simulations, was presented
for the thermodynamic stability of a N$_{\rm{t}}$ fluid when $\kappa$
is at least as large as $7$.
These simulations were supplemented by an extended SPT model that exactly incorporates
the second and third virial coefficients while resumming the remainder of the
virial series. The inclusion of the third virial coefficient increases the
interval in $\kappa$ where the N$_{\rm{t}}$ phase is stable, approaching the
simulation result. Specifically, the I-N$_{\rm{t}}$ transition line moves to lower 
packing fractions, while the intersection point between the I-N$_{\rm{u}}$ 
and the I-N$_{\rm{t}}$ transitions shifts to $\kappa=3.23$. However, this 
value is still lower than the value indicated by simulations.

The properties of the HR fluid are to be contrasted with those of hard discorectangles.
MC simulations of this system have been conducted \cite{Frenkel2}, and the global
phase diagram was computed. A careful inspection of particle 
configurations in the isotropic phase shows that there are peculiar equilibrium 
textures, with large clusters containing particles arranged side by side, exactly
as in the HR fluid. These configurations are favoured by the particular shape of the
particles and by the reduced dimensionality. However, in contrast with the HR fluid, 
neighbouring clusters do not exhibit strong tetratic correlations and, therefore, the 
formation of a tetratic nematic phase is discouraged in the hard discorectangle fluid.

In this article we address the problem of how the present theoretical understanding 
of the HR fluid can be improved by consideration of clustering effects. 
Our thesis is that these effects, very apparent in our own MC simulations but
not addressed by the theories proposed up to now,
are a key factor in the stabilisation of the N$_{\rm t}$ phase. Inclusion
of cluster formation is responsible for the enhancement of the region of N$_{\rm{t}}$ 
stability in the phase diagram. In the model proposed, clustering is approximately taken care of by
treating clusters as distinct species in a mixture of polydispersed rectangles.
The functional form for the cluster size distribution is assumed to be exponential,
an assumption supported by cluster statistical results based on MC simulations and
by a simple chemical--reaction model (see Appendix), and is introduced in the model as an input. 
The thermodynamics of the polydispersed
mixture is analysed using SPT. The results indicate that clustering (assimilated 
in the theory by means of a polydispersity parameter) stabilises the N$_{\rm{t}}$ 
phase for values of aspect ratio much higher than $\kappa =2.62$ if the
polydispersity is sufficiently high. Polydispersity parameters obtained from 
simulation give support to the model. 
    
The article is organized as follows. In Section II we present numerical evidence
that the size distribution in the HR fluid is an exponentially decaying function. 
Section III presents the main ideas of the theoretical model proposed.
The conclusions are drawn in Section IV. 
Finally, details on the model and on the procedure of solution are relegated to 
Appendices A and B, while
Appendix C contains a chemical--reaction model for monomer aggregation which
also supports the assumption of an exponential cluster size distribution.

\section{Monte Carlo simulation of clustering}
\label{II}

We started by applying standard isobaric (NPT) MC techniques on a two--dimensional fluid 
of hard rectangles, with aspect ratios $\kappa=3$, $5$ and $7$, using $N=1400$
rectangles. The transition from the I phase to the N$_{\rm t}$ phase was identified
approximately by inspection of the tetratic order parameter 
$q_2=\left<\cos{4\phi}\right>$, where $\phi$ is the angle between the long axis of
the particle and an axis fixed in space. Once the samples were equilibrated, cluster
statistics was applied. The criterion for pair
connectedness, i.e. for deciding when 
two neighbouring rectangles can be considered to be `bonded', was based on the
relative angle $\phi_{12}$ between the long axes of the particles and their relative
centre--of--mass distance ${\bf r}_{12}$, by demanding that $\phi_{12}<\delta$ and
$|{\bf r}_{12}|<\epsilon$. Typical values adopted were $\delta=10^{\circ}$ and
$\epsilon=1.3 \sigma_0$, although the conclusions to be presented below do not seem
to depend qualitatively on the exact values (provided they are not too large).

\begin{figure}
\epsfig{file=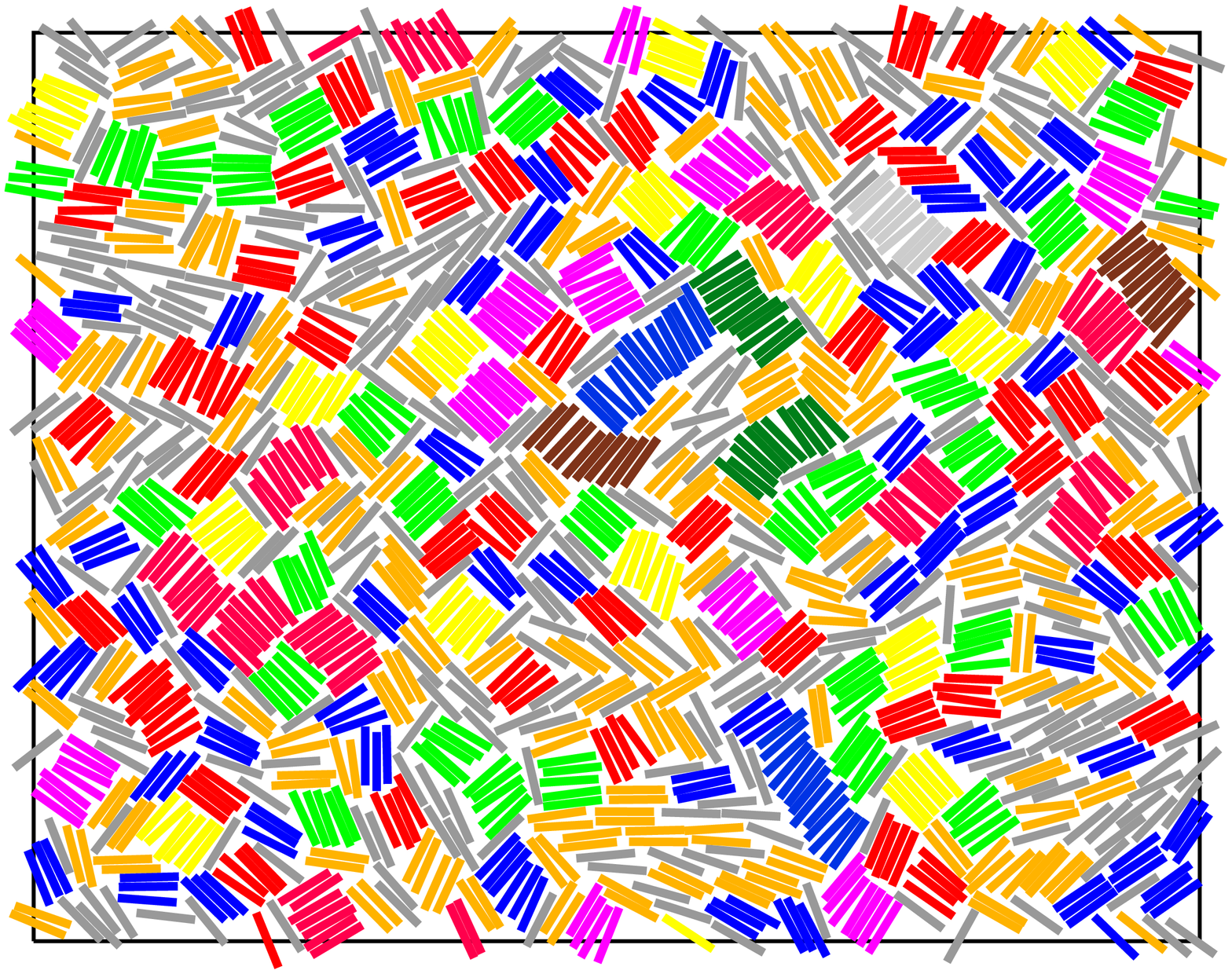,width=3.in}
\caption{
(Colour online). Configuration of hard rectangles of aspect ratio $\kappa=7$
at packing fraction $\eta=0.655$, as obtained from MC simulation.
Clusters, defined by the pair connectedness criterion explained
in the text, have been coloured according to their size.}
\label{MC}
\end{figure}

\begin{figure}
\epsfig{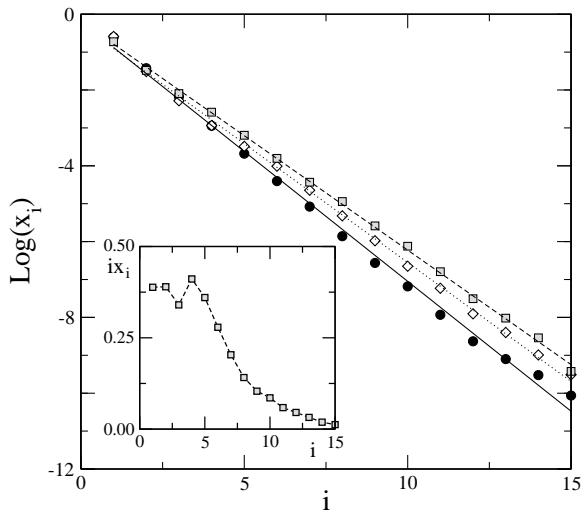}
\caption{
Size distribution function $x_i$ as obtained from MC simulation, 
in logarithmic scale, for 
a fluid of HRs with different values of aspect ratio and packing 
fraction: filled circles, $\kappa=3$ and $\eta=0.635$; open
squares, $\kappa=7$ and $\eta=0.600$; and grey squares, $\kappa=5$ and
$\eta=0.657$. Straight lines are linear fits. Inset: area distribution
function $i x_i$ for the case $\kappa=5$ and $\eta=0.692$.}
\label{MC1}
\end{figure}

Fig. \ref{MC} presents a configuration of rectangles with $\kappa=7$ at a packing fraction
$\eta=0.655$. This corresponds to a tetratic phase. Identified clusters have been 
coloured according to their size. One can see how these clusters look like large
``super--rectangles'' arranged along two perpendicular directions. Therefore, the tetratic
structure is maintained not only for single rectangles, but also at the
level of clusters (``polydispersed super--rectangles''). This hierarchical feature of the
tetratic symmetry will give support to the theoretical model to be presented in the 
following section. In Fig. \ref{MC1} a logarithmic histogram of the size distribution 
(averaged over configurations) is shown; in the figure, $x_i$ is defined as the
fraction of clusters of size $i$ (see next section). The
distribution looks exponential. All the curves in the figure pertain 
to a tetratic phase, but the same behaviour is observed also in the 
isotropic phase (with the general trend that, for given $\kappa$,
the slope decreases, i.e. the size distribution function becomes 
wider, hence the average cluster size, as density increases).
In the size region corresponding to square clusters
(i.e. clusters with aggregation number $i\sim\kappa$) it is possible to see an incipient
shoulder that grows as density is increased. This feature is more apparent in the
area distribution function, $ix_i$ (giving the fraction of area occupied by clusters of
a given size), which presents a shoulder and sometimes even a well--defined peak
(inset of Fig. \ref{MC1}), indicating that square clusters are structurally very relevant 
and contribute very decisively to the thermodynamic properties of the fluid.

\begin{figure}
\epsfig{file=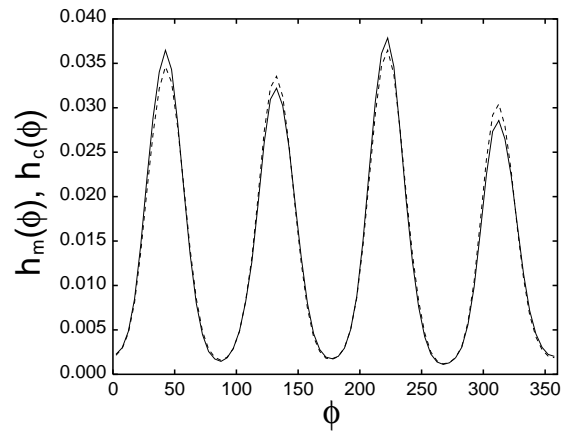,width=3.in}
\caption{  
Monomer $h_m(\phi)$ (solid line) and cluster $h_c(\phi)$  
(dashed line) orientational
distribution functions for the case $\kappa=7$ and $\eta=0.655$, as
obtained from MC simulation.}
\label{MCdist1}
\end{figure}

We end this section by discussing the orientational distribution functions.
Let $h_m(\phi)$ be the monomer orientational distribution function, 
giving the probability of finding a given rectangle with its long axis
forming an angle $\phi$ with respect to the director. Having defined 
clusters in the fluid, we may also define an orientational distribution function
associated with clusters, $h_c(\phi)$, giving the probability of finding 
an average cluster (regardless of its size) oriented with angle $\phi$.
In the simulations we compute $h_c(\phi)$ by averaging over all the identified
clusters, and then over all MC configurations. Fig. \ref{MCdist1} gives
the functions $h_m(\phi)$ and $h_c(\phi)$ for a state with tetratic symmetry.
It is interesting to note that these functions are almost identical (with 
the same
symmetry and, consequently, with peaks with the same height), so that the structure
of tetratic ordering is maintained from the monomer level to the cluster level,
the functions $h_m(\phi)$ and $h_c(\phi)$ obeying a kind of ``similarity''
property. This property does not seem to be followed in the uniaxial nematic
phase.

\section{Theoretical Model} 
\label{III}

The model we propose accounts for clustering effects in an approximate way.
The clear identification of clusters in the simulations, along with their
approximate rectangular shape, leads to a simple model where clusters are
regarded as single rectangular particles with no internal degrees of freedom. Therefore,
we consider a $n$--component mixture of two-dimensional hard rectangles of
dimensions $L$ and $\sigma_i$, with $\sigma_i=i\sigma_0$ and $i$ the
aggregation number. Each of these rectangles is assumed to be composed of
$i$ monomers in perfect contact in a side-by-side configuration.
The total number of monomers, $N_0$, can be written as
\begin{eqnarray}
N_0=\sum_{i=1}^n iN_i, \label{uno}
\end{eqnarray}
where $N_i$ is the number of clusters containing $i$ monomers. 
Dividing by the volume $V$,
\begin{eqnarray}
\rho_0=\sum_{i=1}^ni\rho_i,\label{dos}
\end{eqnarray} 
where $\rho_0=N_0/V$ is the total monomer density, and
$\rho_i=N_i/V=x_i\rho$ is the density of clusters of size $i$, with
$x_i=N_i/N$ their number fraction, while $N=\sum_iN_i$ is the total number
of clusters and $\rho=N/V$ their density. The set $\{x_i\}$, $i=1,\cdots,n$, 
is a central quantity in our model, since it contains information about clustering
tendencies. We will assume $x_i$ to be an exponential with $i$, as explained
later. The total packing fraction of the system is
\begin{eqnarray}
\eta=\sum_{i=1}^n\rho_i a_i,
\end{eqnarray}
where $a_i=ia_0$ ($a_0=L\sigma_0$) is the area of a cluster of size $i$. 
Using (\ref{dos}), we easily obtain $\eta=\rho_0 a_0$.

\begin{figure}
\epsfig{file=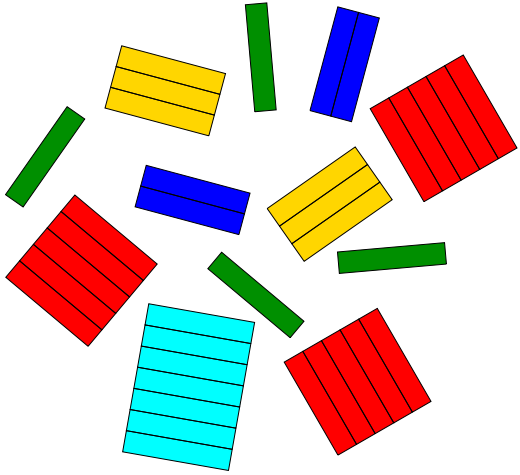,width=3.in}
\caption{
(Colour online).
Multicomponent isotropic mixture composed of different species, where
species, depicted in different colours, correspond to clusters of a particular size. 
Clusters are defined as aggregates of rectangular monomers in a side-by-side configuration.
In this instance the monomer aspect ratio is $\kappa=5$.} 
\label{fig1}
\end{figure}

Fig. \ref{fig1} shows a schematic representation of an isotropic configuration 
of clusters of different sizes (with monomer aspect ratio $\kappa=5$).
The original monomers that give rise to each cluster are indicated but note that,
in our model, the identity of the monomers is lost, as monomers in the same cluster
do not interact dynamically, always being in perfect side-by-side contact. 

The free--energy of this multicomponent mixture of hard rectangles will be modelled
by means of SPT applied to a fluid mixture of freely--rotating hard rectangles. The
orientational properties of the mixtures will be characterised by orientational
distribution functions $h_i(\phi)$ for each component $i$. The free--energy density 
functional $\Phi=\beta F/V$ is written as
\begin{eqnarray}
\Phi[\{h_i\}]&=&\rho\Bigg\{\sum_{i=1}^nx_i\left[\ln 
\left(x_i{\cal V}_i\right)+\int_0^{\pi}d\phi h_i(\phi)
\ln[h_i(\phi)\pi]\right]\nonumber\\
&-&1+\ln y+y S\left[\{h_i\}\right]\Bigg\},
\label{Phi}
\end{eqnarray}
where ${\cal V}_i$ is the thermal volume of $i$--sized clusters, and 
we defined $y=\rho a_0/(1-\eta)$. Due to the head--tail symmetry of the
particles, the angle 
$\phi$ can be restricted to the interval $[0,\pi]$ and the functions 
$h_i(\phi)$ normalised accordingly. The function
$S\left[\{h_i\}\right]=\sum_{i,j}x_ix_j S_{ij}\left[\{h_i\}\right]$, with 
\begin{eqnarray}
S_{ij}\left[\{h_i\}\right]&=&\frac{1}{2}\left(\kappa+ij\kappa^{-1}\right)
\langle\langle|\sin\phi_{ij}|\rangle\rangle\nonumber\\
&+&
\frac{1}{2}(i+j)\langle\langle|\cos\phi_{ij}|\rangle\rangle,
\label{Sij}
\end{eqnarray}
is related to $A_{ij}$, 
the angle--averaged excluded area between clusters $i$ and $j$, as 
$S_{ij}=(A_{ij}/a_0-i-j)/2$. The shorthand notation 
$\langle\langle f(\phi_{ij})\rangle\rangle$ has been used
for the double angular average of a generic function:
$\langle\langle f(\phi_{ij})\rangle\rangle=\int_0^{\pi} d\phi_i\int_0^
{\pi} d\phi_j 
h_i(\phi_i)h_j(\phi_j)f(\phi_{ij})$.
Now a bifurcation analysis of (\ref{Phi}) 
at the I-N$_{u,t}$ transition (see Appendix) allows 
us to obtain the packing fractions of the I-N$_{u,t}$ spinodal lines as
\begin{eqnarray}
\eta^*=\left[1-\frac{4}{\pi}g_k\left(\frac{\kappa}{m_0^{(1)}}+
\frac{m_0^{(2)}}{\kappa m_0^{(1)}}+2(-1)^k\right)\right]^{-1},  
\label{spino}
\end{eqnarray}
with $k=1$ for the uniaxial and $k=2$ for the tetratic nematic, while 
$m_0^{(\alpha)}=\sum_i x_i i^{\alpha}$ ($\alpha=1,2$), are the 
first and second moments of the cluster size distribution function.

Based on the MC results, we adopt an exponential cluster size distribution:
\begin{eqnarray}
x_i=\frac{1-q}{1-q^n}q^{i-1},\hspace{0.4cm}i=1,\cdots,n,
\label{exp}
\end{eqnarray}
with $q=e^{-\lambda}$ ($\lambda>0$). The prefactor in (\ref{exp}) ensures 
that the distribution is normalised, i.e. that $\sum_i x_i=1$. 
The first two moments can be derived analytically:
\begin{eqnarray}
m_0^{(1)}&=&\frac{1-\left[1+n(1-q)\right]q^n}{(1-q)(1-q^n)},\label{m1}\\
m_0^{(2)}&=&\frac{1+q-\left[q+\left(1+n(1-q)\right)^2\right]q^n}{
\label{m2}
(1-q)^2(1-q^n)}.
\end{eqnarray}

\begin{figure}
\epsfig{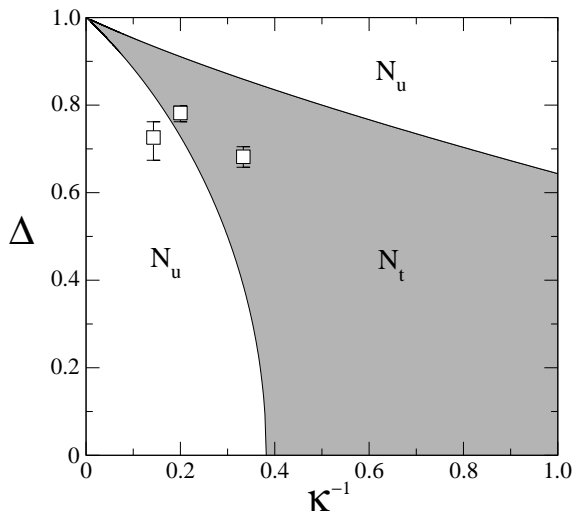}
\caption{
The functions $\Delta(\kappa)$ in terms of inverse aspect ratio
$\kappa^{-1}$. Symbols indicate the nematic phase which is 
stable in the region in
question. The region of tetratic stability is shaded.
Open squares indicate values of polydispersity, for values of 
aspect ratio $\kappa=3$, $5$ and $7$, at the I--N$_{\rm t}$ transition,
as obtained from MC simulation; error bars correspond to uncertainty
in $\Delta$ originating from uncertainty in location of spinodal.} 
\label{dos0}
\end{figure}

Now we use (\ref{spino}) to obtain the maximum aspect ratio which can support a
stable tetratic phase. This follows by imposing the condition $\eta^*_{0,t}=
\eta^*_{0,u}$, i.e. by searching for the intersection point of the two
spinodal lines I--N$_{\rm u}$ and I--N$_{\rm t}$ in the phase diagram $\eta-\kappa$.
Solving for the corresponding value of $\kappa$, we obtain
\begin{eqnarray}
\kappa=\frac{1}{2}\left(3m_0^{(1)}\pm\sqrt{9\left(m_0^{(1)}\right)^2-
m_0^{(2)}}\right).
\label{kap}
\end{eqnarray}
In the specific case where the number of species goes to infinity, $n\to\infty$,
we obtain from (\ref{m1})-(\ref{kap}) 
\begin{eqnarray}
\Delta=\kappa^{-1}\left[\frac{1}{2}\left(2\kappa^2-3\kappa-1\pm\sqrt{\kappa^2
+6\kappa+1}\right)\right]^{1/2},
\label{11}
\end{eqnarray}
where $\Delta=\sqrt{m_0^{(2)}/\left(m_0^{(1)}\right)^2-1}=\sqrt{q}$, 
a measure of polydispersity, is the relative mean square deviation. 
The two functions $\Delta(\kappa)$, corresponding to the two signs in
(\ref{11}), are plotted in Fig. \ref{dos0} (note that one of
the branches, the one with minus sign, can be only calculated for
$\kappa\geq \kappa^*=(3+\sqrt{5})/2$). The two lines define a region (shaded in the
figure) where the tetratic phase may be stabilised. Thus we see how, as the value of
polydispersity $\Delta$ is increased, the maximum value of $\kappa$ for which the 
tetratic phase ceases to exist increases, which means that polydispersity
enhances the formation of tetratic ordering. 

Note that the values of $\Delta$ at the upper boundary of the tetratic 
region are very high, which means that there should be a large number of clusters
with their long side $i\sigma_0\gg L$, while their width is $L$. This causes
the effective average cluster aspect ratio $i\sigma_0/L$ to be very high, 
which induces formation of {\it uniaxial} nematic order and thus the N$_{\rm t}$
phase is destabilised in favour of the N$_{\rm u}$ phase. The presence of such
big clusters is not observed in the simulations, which means that a more
sophisticated model should somehow include additional entropy terms that
discourage the formation of big clusters; for the sake of simplicity, we have 
kept the ingredients of our model to a minimum and avoided any such additional
complications.

Finally, in Fig. \ref{dos0} values of polydispersity $\Delta$ calculated from 
the simulated cluster size distribution function have been indicated by symbols.
These values correspond to the estimated spinodal line of the I--N$_{\rm t}$
transition for the cases $\kappa=3$, $5$ and $7$. Since these estimations
are very rough, the uncertainty in packing fraction at the spinodals 
goes over to the value of polydispersity (which naturally depends on
$\kappa$ and $\eta$), which is represented in each case
by error bars. We can see that in two of the cases the symbols are
well inside the tetratic stability region calculated from the theory.
In the case $\kappa=7$, where uncertainty is larger, 
the MC estimate of $\Delta$ lies outside (but close to) this region.

\section{Conclusions}

Because of the reduced dimensionality, fluids of two--dimensional
hard anisotropic particles 
exhibit strong clustering effects: particles have less freedom to orient in
space, which fosters configurations where neighbouring particles lie parallel
to each other. However, the impact of this on the onset of new macroscopic
symmetries depends very sensitively on the particular geometry of the 
particles. Thus, in fluids of rectangles, neighbour clusters have a strong
tendency to adopt orthogonal relative configurations, since these clusters
are almost perfect {\it big} rectangles made of several, almost parallel, 
monomers. These strong tetratic correlations are capable of generating full
macroscopic tetratic order and a thermodynamically stable tetratic phase.
Therefore, clustering effects are crucial to understand phase behaviour
in the HR fluid (and possibly also in the HDR and related fluids), but
simple theories at the level of two--body {\it monomer} 
correlations (Onsager, SPT, etc.) cannot account for these effects. 

In this work we have presented a simple
theory that incorporates clustering in terms of cluster polydispersity, where
clusters are considered to be inert particles with no internal degrees of
freedom. This assumption may be accurate provided the cluster lifetime
(i.e. the average time it takes for a cluster to disappear since it was
formed) is longer than the typical cluster diffusion rates in the fluid. 
Validation of this condition will have to wait for molecular dynamics simulations
of the hard--rectangle fluid. Once the fluid is modelled in terms of a
multicomponent mixture, one of the available theories for mesophase
formation can be used. We have used SPT and have examined the consequence
of polydispersity in the phase diagram. As expected, polydispersity enhances
the stability of the tetratic phase. Due to the limitations of our model
(e.g. the cluster size distribution has to be imposed from outside and
does not result from the theory), we cannot make any quantitative 
comparison with available simulation and experimental results. However,
the model can qualitatively explain the formation of tetratic order for 
rather high values of aspect ratio, as shown by simulation and experiment.

\acknowledgments

Y.M.-R. gratefully acknowledges financial support from Ministerio
de Educaci\'on y Ciencia (Spain) under a Ram\'on y Cajal research
contract and the MOSAICO grant. This work has been partly financed by 
grants Nos. FIS2005-05243-C02-01 and FIS2007-65869-C03-01, also from
Ministerio de Educaci\'on y Ciencia, and S-0505/ESP-0299 from
Comunidad Aut\'onoma de Madrid (Spain).

\section{Appendix}

In this appendix we provide additional details and further information 
on the consequences of the model. It contains two sections. In Section A,
details on the set of non--linear equations that have to be solved to obtain
the equilibrium properties of the HR fluid are provided. Also, the
bifurcation analysis of the I--N$_{\rm{u,t}}$ transitions is presented.
Section B is devoted to discussing the nature of the different phase transitions, 
together with the behaviour of the distribution functions and to a 
comparison with simulations. In Section C, a simple chemical model of aggregation
is discussed.

\subsection{Minimisation of free energy and bifurcation analysis}

Using Fourier series to represent the orientational distribution functions,
\begin{eqnarray}
h_i(\phi)=\frac{1}{\pi}\sum_{k\geq 0}h_k^{(i)}\cos(2k\phi),
\end{eqnarray}
with $h_0^{(i)}=1$ $\forall i$, together with Eqn. (\ref{Sij}), we find 
\begin{eqnarray}
S_{ij}=\frac{1}{\pi}\sum_{k\geq 0}\left[\kappa+\frac{ij}{\kappa}+
(-1)^k(i+j)\right]g_kh_k^{(i)}h_k^{(j)},
\end{eqnarray}
where $g_k=-(1+\delta_{k0})/2(4k^2-1)$. Defining
\begin{eqnarray}
m_k^{(\alpha)}=\sum_i x_i i^{\alpha}h_k^{(i)},\quad \alpha=0,1,
\end{eqnarray}
we obtain 
\begin{eqnarray}
\sum_{ij}x_ix_jS_{ij}=\frac{\kappa}{\pi}\sum_k 
g_k s_k^2,\hspace{0.4cm}
s_k=m_k^{(0)}+(-1)^k\frac{m_k^{(1)}}{\kappa}.\nonumber \\
\end{eqnarray}
Note that $m_0^{(0)}=\sum_i x_i=1$ while $m_0^{(1)}=\sum_i x_i i$ is the first 
moment of the discrete cluster size distribution function $\{x_i\}$. 
Using this notation, the free-energy per particle $\varphi=\Phi/\rho_0$ can be written
\begin{eqnarray}
&\varphi&=\ln\left(\frac{y_0}{m_0^{(1)}}\right)-1+
\sum_{i=1}^nx_i\Bigg\{\ln\left(x_i{\cal V}_i\right)\nonumber\\
&+&\int_0^{\pi}
d\phi h_i(\phi)\ln\left[h_i(\phi)\pi\right]\Bigg\}+
\frac{y_0\kappa}{\pi m_0^{(1)}}\sum_kg_ks_k^2,
\label{varphi}
\end{eqnarray}
with $y_0=\eta/(1-\eta)$. The functional minimization of (\ref{varphi}) 
with respect to $h_i(\phi)$ gives a set of self--consistent non-linear 
equations which, after some algebraic manipulations, can be transformed into 
a set of equations for the new variables $s_k$:
\begin{eqnarray}
&&s_k=2\sum_{i=1}^nx_i\left[1+\frac{(-1)^k}{\kappa}i\right]
Q_k^{(i)},
\label{nolinear}\\
&&Q_k^{(i)}=\int_0^{\pi}d\phi\cos(2k\phi)h_i(\phi),
\end{eqnarray}  
where the normalized orientational distribution functions are
\begin{eqnarray}
&&h_i(\phi)=\frac{\displaystyle e^{-\Lambda_i(\phi)}}
{\displaystyle\int_0^{\pi}d\phi e^{-\Lambda_i(\phi)}}, \\ 
&&\Lambda_i(\phi)=\frac{4y_0\kappa}{\pi m_0^{(1)}}\sum_{k\geq 1} 
s_k\left[1+\frac{(-1)^k}{\kappa}i\right]g_k\cos(2k\phi).\nonumber\\
\label{nolinear1}
\end{eqnarray}
The linearization of (\ref{nolinear}) with respect to $s_k$ ($k\geq 1$) 
allow us to obtain the expression (\ref{spino})
for the packing fractions at the I-N$_{u,t}$ spinodal lines. 

\subsection{Phase transitions and distribution functions}

\begin{figure}
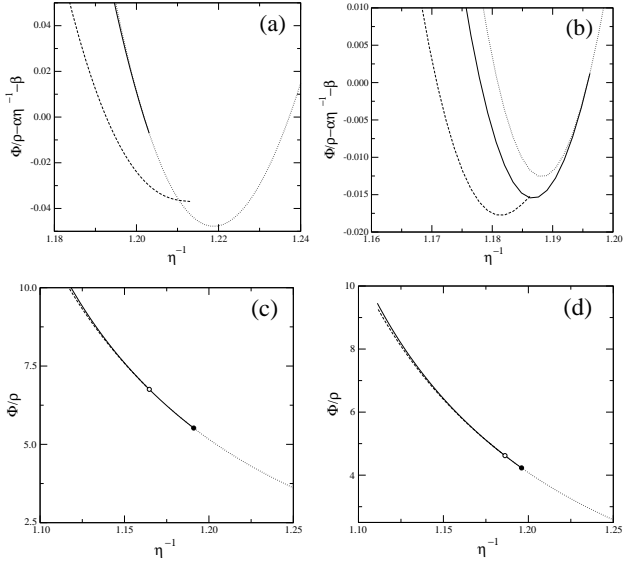

\epsfig{file=fig6a.eps,width=1.6in}
\epsfig{file=fig6b.eps,width=1.6in}
\epsfig{file=fig6c.eps,width=1.55in}
\hspace*{0.1cm}
\epsfig{file=fig6d.eps,width=1.55in}
\caption{
Free energy per particle $\Phi/\rho$ vs. inverse
packing fraction $\eta^{-1}$ from SPT results for the
multicomponent HR fluid of monomer
aspect ratio $\kappa=3$ and with different values of polydispersity: 
(a) $q=0.25$, (b) $0.35$, (c) $0.50$ and (d) $0.65$. 
Continuous curves: tetratic phase; dashed curves: uniaxial nematic
phase; dotted curves: isotropic phase. In (c) and (d) symbols indicate
bifurcation points from the isotropic to the tetratic (filled circles)
and uniaxial nematic (open circles) phases. In (a) and (b) a straight 
line in $\eta^{-1}$ has been subtracted to better visualise the 
curves.}
\label{monton}
\end{figure}

In this subsection we analyse the free energy branches of the model
in order to understand the nature of the different phase transitions.
The cluster distribution function $x_i$ of the mixture is assumed
to be exponential, and the width of the distribution is fixed via
the polydispersity parameter $q$ (or $\Delta$). Eqns.  
(\ref{nolinear})--(\ref{nolinear1}) are solved for different values of $\eta$ 
to find all metastable and stable phases, either I, N$_{\rm{u}}$ or 
N$_{\rm{t}}$ phases. In Fig. \ref{monton} the free-energy branches as a function 
of $\eta^{-1}$ for different values of $q$ (and hence for different 
polydispersities) are shown. 
As can be seen, the N$_{\rm t}$ phase begins to be stable from
$q\approx 0.30$. Also, it is clear that, for $q=0.25$ and $0.35$,
the I-N$_{\rm{u}}$ or N$_{\rm{t}}$-N$_{\rm{u}}$ transitions are 
of first order (free--energy branches cross with different slopes). 
The coexistence values of $\eta$ cannot be determined from 
the standard double--tangent construction, since the present system is 
polydisperse. The usual procedure then is 
to fix the distribution function $x_i^{(0)}$ and packing fraction $\eta^{(0)}$ 
for the parent phase (I or N$_{\rm{u,t}}$ phases) and find the cloud 
and shadow curves. We have not implemented this procedure here. 
However, for those values of $q$ for which the transitions are continuous 
(i.e. $q=0.5$ and $0.65$), the present procedure adequately determines 
the transition densities. A similar situation occurs for the cases
$\kappa=5$ and $\kappa=7$ (not shown). 

\begin{figure}
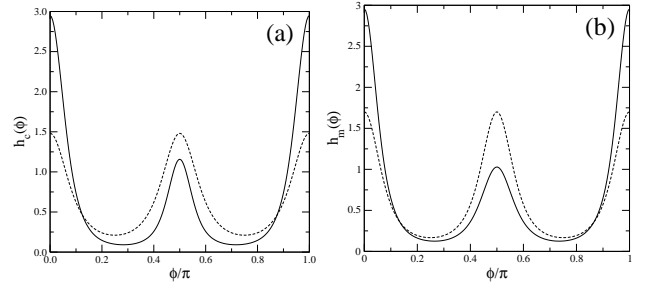

\epsfig{file=fig7a.eps,width=1.6in}
\epsfig{file=fig7b.eps,width=1.6in}
\caption{
Cluster (a) and monomer (b) orientational distribution functions 
for $\kappa=5$ and $q=0.65$. The continuous line corresponds to
a stable N$_{\rm u}$ phase with $\eta=0.85$, whereas the discontinuous line is for
the metastable N$_{\rm t}$ phase at the same value of $\eta$.}
\label{monton4}
\end{figure}

It is also interesting to look at the orientational distribution functions
of monomers and clusters. From the corresponding functions for clusters
of size $i$, i.e. $h_i(\phi)$, it is easy to define a cluster 
orientational distribution function $h_c(\phi)$ as
\begin{eqnarray}
h_c(\phi)=\sum_{i=1}^n x_i h_i(\phi).
\end{eqnarray}
From this, order parameters of the multicomponent mixture can also be
defined:
\begin{eqnarray}
Q^{(k)}=\sum_i x_i Q_i^{(k)},\quad k=1,2.
\end{eqnarray}
In the case of monomers the situation is a bit more complicated, since
in our model we have lost track of monomers as distinct entities. However,
we can simply count the number of monomers pointing along some angle $\phi$
from the set $h_i(\phi)$ and then divide by the average number of clusters.
Here we have to bear in mind that clusters of size $i$ (having $i$ monomers)
with $i<\kappa$ (and $\kappa$ an integer)
have a long axis in a direction perpendicular to that of
clusters with $i>\kappa$; therefore we write:
\begin{eqnarray}
h_m(\phi)=\frac{1}{m_0^{(1)}}\left[\sum_{i=1}^{n_0} x_i i h_i(\phi)+
\sum_{i=n_0+1}^n x_i i h_i(\phi+\pi/2)\right],\nonumber\\
\end{eqnarray}
with $n_0=[\kappa]$ (note that in the case where $\kappa$ is not an integer 
this division has to be done also).
Thus the order parameters of the monomers can be calculated as
\begin{eqnarray}
\langle \cos(2\phi)\rangle_m&=&\frac{1}{m_0^{(1)}}\Bigg|
\sum_{i=1}^{n_0}x_i iQ_i^{(1)}-
\sum_{i=n_0+1}^nx_i iQ_i^{(1)}\Bigg|,\nonumber \\ \\
\langle \cos(4\phi)\rangle_m&=&
\frac{1}{m_0^{(1)}}\sum_{i=1}^{n_0}x_i iQ_i^{(2)}.
\end{eqnarray}
Even for values of aspect ratio for which there exists
a region of tetratic stability, the N$_{\rm{u}}$ is always the more 
stable phase for high values of packing fraction. 
It is interesting to note that, in this situation, the cluster and monomer
distribution functions $h_{\rm{c,m}}(\phi)$ usually have a secondary 
peak at $\phi=\pi/2$ corresponding to tetratic ordering (this feature
is also present in the simple SPT for the one--component HR fluid). Fig.
\ref{monton4} shows these distributions for the case $\kappa=5$ and
$q=0.65$. Note that, according to Fig. \ref{dos0}, there exists a 
stable tetratic phase in this case, with a value of packing fraction
at the N$_{\rm t}$--N$_{\rm u}$ transition of $\eta^*=0.839$. In the
figure, the packing fraction chosen is $\eta=0.85 > \eta^*$ and, 
therefore, the stable phase has an orientational 
distribution function pertaining to a N$_{\rm{u}}$. 
However, both the cluster and the monomer orientational distribution 
functions of N$_{\rm{u}}$ have secondary peaks. 
We have also 
plotted in Fig. \ref{monton4} the distribution functions of a metastable 
N$_{\rm{t}}$ at the same value of $\eta$.
Finally, it is also interesting that,
although the distribution functions $h_m(\phi)$ and $h_c(\phi)$ in the
tetratic region are always tetratic--like (i.e. all maxima have the same 
height, as it should be by construction), they do not coincide in the uniaxial
nematic phase and, consequently, the similarity property at work in the tetratic
phase is not obeyed for the uniaxial nematic (see Section \ref{II}).

\subsection{Chemical reaction model}

Aggregation phenomena in dilute fluids (e.g. micelle aggregation)
are very often described in terms of a chemical reaction model.
An exponentially decaying size distribution immediately emerges
from these models. Here we exploit the idea and carry it further
using our density--functional approximation. This model assumes
that the lifetime of a cluster is sufficiently long that it can
be defined as a distinct `chemical' species.

One assumes a chemical reaction of the type $C_l+C_1\rightleftharpoons 
C_{l+1}$, with $C_l$ denoting a `chemical' species (i.e. a cluster)
containing $l$ monomers. Chemical equilibrium between clusters and 
monomers then implies the relation
\begin{eqnarray}
\mu_i=i\mu_1,\quad i=2,\cdots,n. \label{cuatro}
\end{eqnarray}
Now the chemical potential of the $i$--th species can be calculated from
our density--functional theory as
\begin{eqnarray}
\beta \mu_i=\frac{\partial\Phi}{\partial\rho_i},
\end{eqnarray}
which results in
\begin{eqnarray}
&&\beta \mu_i=\ln\left( x_i{\cal V}_i\right) +\int_0^{\pi}d\phi h_i(\phi) 
\ln[h_i(\phi)\pi]+\ln y\nonumber\\
&&+y i +2y\sum_j x_j S_{ij}+y^2S i.
\label{chepo}
\end{eqnarray}
The $n-1$ Eqns. (\ref{cuatro}), together with the condition (\ref{dos}),
are a set of $n$ equations with $n$ unknowns ($x_2,\cdots, x_n$ and $\rho$) 
which allow to find the equilibrium configuration of the fluid. Due to the
simplicity of the model an analytical solution can be found.

\begin{figure}
\epsfig{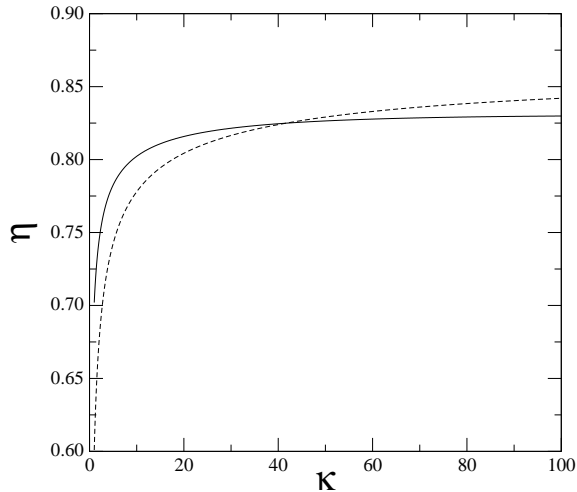}
\caption{
Packing fractions $\eta$ of the I--N$_{\rm t}$ (continuous line)
and I--N$_{\rm u}$ (dashed line) transitions versus aspect ratio $\kappa$.} 
\label{figa}
\end{figure}
Since, for the isotropic phase, we have 
\begin{eqnarray}
\sum_j x_j\left(iS_{1j}-S_{ij}\right)=\frac{(i-1)}{\pi}(\kappa+m_0^{(1)}),
\end{eqnarray}
Eqns. (\ref{cuatro}), together with (\ref{chepo}), give $x_i=x_1 q^{i-1}$ with
\begin{eqnarray}
q=x_1\frac{y_0}{m_0^{(1)}}\exp\left[\frac{2y_0}{\pi}\left(\frac{\kappa}{m_0^{(1)}}+1
\right)\right].
\label{final}
\end{eqnarray}
Here we assumed that ${\cal V}_i={\cal V}_1^i$ (consistent with the 
absence of internal degrees of freedom in the clusters and also with the
assumption that all clusters have the same mass). 
Now, since $\sum_i x_i=1$, we find,
for a fluid with an infinite number of species,
$1=x_1/(1-q)$ which, together with (\ref{final}), give
\begin{eqnarray}
x_1=\left\{1+\frac{y_0}{m_0^{(1)}}\exp\left[\frac{2y_0}{\pi}
\left(\frac{\kappa}{m_0^{(1)}}+1
\right)\right]\right\}^{-1}.
\end{eqnarray}
Also, the first moment can be calculated self--consistently as
\begin{eqnarray}
&&m_0^{(1)}=x_1\sum_{i=1}^{\infty}i q^{i-1}=\frac{x_1}{(1-q)^2}\nonumber\\&&=
1+\frac{y_0}{m_0^{(1)}}\exp\left[\frac{2y_0}{\pi}\left(\frac{\kappa}{m_0^{(1)}}+1
\label{mom1}
\right)\right].
\end{eqnarray}
This solution means that 
$x_i=\left(1-1/m_0^{(1)}\right)^{i-1}/m_0^{(1)}$.
While the first moment $m_0^{(1)}$ is the solution of 
Eqn. (\ref{mom1}), the second moment results in
\begin{eqnarray}
m_0^{(2)}=x_1\sum_{i=1}^{\infty}i^2q^{i-1}=x_1\frac{1+q}{(1-q)^3}=
m_0^{(1)}(2m_0^{(1)}-1).\nonumber\\
\end{eqnarray}
The polydispersity coefficient, defined as $\Delta=\sqrt{m_0^{(2)}/
\left(m_0^{(1)}\right)^2-1}$, turns out to be
$\Delta=\sqrt{1-1/m_0^{(1)}}$.

\begin{figure}
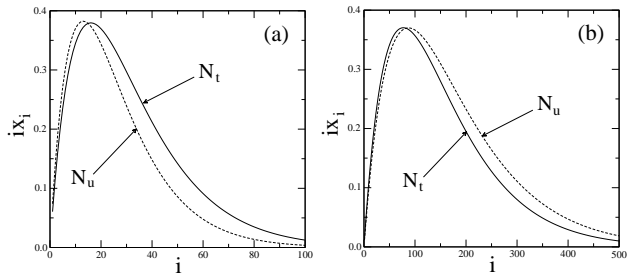

\epsfig{file=fig9a.eps,width=1.6in}
\epsfig{file=fig9b.eps,width=1.6in}
\caption{
Area distribution functions for tetratic N$_{\rm t}$ and
uniaxial N$_{\rm u}$ nematic phases. (a) $\kappa=10$ and (b)
$\kappa=100$.} 
\label{cuatro0}
\end{figure}

Now the I-N$_u$ and I-N$_t$ spinodals can be calculated by solving 
Eqn. (\ref{mom1}), together with the value of $y_0$ obtained from 
Eq. (\ref{spino}). 
Fig. \ref{figa} contains the functions $\eta_{u,t}(\kappa)$ 
obtained as the solutions of (\ref{mom1}) and (\ref{spino}). As 
can be seen the uniaxial nematic is more stable than the tetratic up 
to $\kappa^*\approx 41.65$. This results from the peculiar behaviour of
the area distribution function $i x_i$: its zeroth--order moment is always 
greater than $\kappa$, while it is larger in the tetratic phase when 
$\kappa<\kappa^*$ [Fig. \ref{cuatro0}(a)]; this behaviour is inverted for 
$\kappa>\kappa^*$, and the moment becomes larger for the uniaxial nematic
phase, as can be seen in Fig. \ref{cuatro0}(b). Again this peculiar 
behaviour is due to the relatively
high proportion of very big clusters, with $i\sigma_0>L$,
which stabilise the N$_{\rm u}$ phase against the N$_{\rm t}$ phase. This
behaviour is not observed in simulations, because the formation of very big
clusters is penalised by fluctuations. The model has the value that 
an exponential cluster distribution function is predicted.

\end{document}